\documentclass[12pt]{article}

\usepackage{amssymb}
\usepackage{amsmath}
\usepackage{amscd}
\usepackage{latexsym}
\usepackage{graphicx}

\usepackage{cite}

\newcommand{\be}{\begin{equation}}
\newcommand{\ee}{\end{equation}}

\newcommand{\bB}{{\bf B}}
\newcommand{\bS}{{\bf S}}
\newcommand{\bt}{\beta}

\newcommand{\al}{\alpha}

\newcommand{\gm}{\gamma}
\newcommand{\om}{\omega}

\newcommand{\rgl}{\rangle}
\newcommand{\lgl}{\langle}

\begin{document}

\begin{center}

{\Large{\bf Regulating spin dynamics of graphene flakes} \\ [5mm]

V.I. Yukalov$^{1,2}$, V.K. Henner$^{3,4}$, and T.S. Belozerova$^4$} \\ [3mm]

{\it
$^1$Bogolubov Laboratory of Theoretical Physics \\
Joint Institute for Nuclear Research, Dubna 141980, Russia \\ [3mm]

$^2$Instituto de Fisica de S\~ao Carlos, Universidade de S\~ao Paulo \\
CP 369, S\~ao Carlos 13560-970, S\~ao Paulo, Brazil \\ [3mm]

$^3$Department of Physics, University of Louisville \\
Louisville, Kentucky 40292, USA  \\ [3mm]

$^4$Department of Physics, Perm State University \\
Perm 614990, Russia}

\end{center}

\vskip 2cm

\begin{abstract}
A method of regulating spin dynamics of the so-called magnetic graphene is
analyzed. Magnetic moments can be incorporated into graphene flakes and graphene
ribbons through defects, such as adatoms and vacancies. Local spins can also be
attached to graphene at hydrogenated zigzag edges. Spin flips can be produced by
transverse pulses and by connecting the sample to a resonance electric coil. The
action of the resonator feedback field strongly accelerates spin reversal. The
possibility of fast spin reversal is important for spintronics and for quantum
information processing allowing for an efficient functioning of spin registers.
\end{abstract}

\vskip 1cm
{\parindent =0pt
{\bf Keywords}:
Spin dynamics; graphene flakes; magnetic defects; magnetization reversal }

\newpage

\section{Traditional approaches}

Graphene is an allotrope of carbon in the form of a sheet of a single layer
(monolayer) of carbon atoms, tightly bound in a hexagonal honeycomb lattice with
a molecular bond length of $0.142$ nanometers. The properties of graphene have
been described in detail in several reviews
\cite{Neto_1,Abergel_2,Dresselhaus_3,Sarma_4,Goerbig_5,Saito_6,Kotov_7,Cann_8,Wehling_9}
discussing a number of applications within different scientific disciplines, e.g.,
in high-frequency electronics, bio sensors, chemical sensors, magnetic sensors,
ultra-wide bandwidth photodetectors, energy storage and generation.

The so-called magnetic graphene can be created by incorporating local defects,
such as adatoms and vacancies, into graphene flakes and ribbons
\cite{Yaziev_10,Palacios_11,Katsnelson_12,Enoki_13}. This can be done in different
ways, for instance, by irradiating graphene flakes or using mechanical and chemical
methods of incorporating defects. Local magnetic moments can be attached to graphene
flakes at hydrogenated zigzag edges, forming a kind of nanomolecules exhibiting
ferromagnetism at room temperatures. The hydrogenated zigzag-edge groups can possess
local magnetic moments with spins $1/2$, $3/2$, and $5/2$.

It is the magnetic properties of graphene that is the topic of the present chapter.
More precisely, our aim is to describe how the direction of magnetic moments could
be regulated. The possibility of regulating the polarization of spins is of principal
importance for spintronics as well as for information processing.

The most often used method of realizing spin reversal is by acting on polarized
spins by an alternating transverse magnetic field. For the efficient spin reversal,
it is necessary to find the appropriate strength of the alternating field, its
frequency, and duration. Below we describe another method in which the required
field, with the necessary features, is created automatically, being a feedback
field.

\section{Magnetic defects coupled to a resonance electric circuit}

Magnetic defects in graphene interact through exchange forces, which can be
characterized \cite{Yaziev_10,Palacios_11,Katsnelson_12,Enoki_13} by the anisotropic
Heisenberg model
\be
\label{1}
 \hat H_{def} = -\; \frac{1}{2} \sum_{i\neq j} J_{ij} \left[ \left(
S_i^x S_j^x + S_i^y S_j^y \right) + \al S_i^z S_j^z \right] \; ,
\ee
in which $S_i^\nu$ is a $\nu$ spin component of the $i$-th magnetic defect,
$J_{ij}$ is the exchange interaction potential, and $\alpha$ is an anisotropy
parameter. The index $i=1,2,\ldots,N$ enumerates the sites of magnetic defects.
The typical values of the exchange interactions for the nearest neighbors are
$J_{ij} < 0.1$ eV, that is $J_{ij}<10^{-13}$ erg. Hence $J_{ij}/\hbar<10^{14}$
s$^{-1}$. The magnetic anisotropy usually is not large, such that $\al$ is close
to one.

Adding the Zeeman part gives us the total Hamiltonian
\be
\label{2}
 \hat H = \hat H_{def} - \mu_S \sum_{j=1}^N \bB \cdot \bS_j \;  ,
\ee
where $\mu_S$ is the magnetic moment of a defect with spin $S$. The external
magnetic field consists of two terms,
\be
\label{3}
  \bB = B_0 {\bf n}_z + H {\bf n}_x \; .
\ee
One term is a constant magnetic field $B_0$ along a unit vector ${\bf n}_z$. In
addition, the sample is assumed to be inserted into a magnetic coil of an electric
circuit, so that moving spins induce electric current in the circuit, which creates
a magnetic feedback field $H$. With the coil axis along the unit vector ${\bf n}_x$,
the feedback field acts along ${\bf n}_x$.

The electric circuit contains resistance $R$, inductance $L$, and capacity $C$. The coil
has $n$ turns, length $l$, and cross-sectional area $A_{coil}$, hence volume
$V_{coil} = A_{coil} l$. The induced current is given by the Kirchhoff equation
\be
\label{K1}
L\; \frac{dj}{dt} + R j + \frac{1}{C} \int_0^t j(t')\;dt' = -\; \frac{d\Phi}{dt} \; ,
\ee
in which $\Phi$ is the magnetic flux
\be
\label{K2}
 \Phi = \frac{4\pi}{c}\; n A_{coil} \eta m_x \;  ,
\ee
where $\eta \approx V/V_{coil}$ is the filling factor, and $V$ is the sample volume. The
magnetic flux is formed by the $x$ component of the moving magnetization density
\be
\label{K3}
 m_x = \frac{\mu_s}{V} \sum_{j=1}^N \; \lgl \; S_j^x \; \rgl \;  .
\ee
The induced electric current, circulating over the coil, creates a magnetic field
\be
\label{K4}
 H \equiv \frac{4\pi n}{cl} \; j \;  .
\ee
The circuit natural frequency is given by the expression
\be
\label{K5}
\om \equiv    \frac{1}{\sqrt{LC} } \qquad
\left( L \equiv 4\pi \; \frac{n^2A_{coil}}{c^2l} \right) \;
\ee
and the circuit damping by
\be
\label{K6}
 \gm \equiv \frac{R}{2L} =  \frac{\om}{2Q} \; ,
\ee
with $Q$ being the quality factor
\be
\label{K7}
 Q \equiv   \frac{\om L}{R} = \frac{\om}{2\gm} \;  .
\ee
Then from the Kirchhoff equation for the induced electric current, we get \cite{Yukalov_14}
the equation for the feedback magnetic field
\be
\label{K8}
 \frac{dH}{dt} + 2\gm H + \om^2 \int_0^t H(t')\;dt' = - 4\pi \eta \; \frac{dm_x}{dt} \; .
\ee

Finally, introducing the dimensionless feedback field
\be
\label{4}
 h \equiv \frac{H}{B_0}
\ee
and differentiating equation (\ref{K8}), we come to the equation
\be
\label{5}
 \frac{d^2h}{dt^2} + \frac{1}{Q} \; \frac{dh}{dt} + h = 4\pi\bt \;
\frac{d^2e_x}{dt^2} \; ,
\ee
in which the parameter
\be
\label{7}
 \bt = \left| \;  \frac{\mu_S SN}{B_0V_{coil}} \right|
\ee
characterizes the strength of coupling of the sample with the circuit. The electromotive
force in the right-hand side of equation (\ref{5}) is formed by moving spins
\be
\label{8}
 e_{\nu j} = \frac{1}{S} \; \lgl \; S_j^\nu \; \rgl \qquad ( \nu = x,y,z ) \; ,
\ee
with
\be
\label{9}
e_\nu = \frac{1}{N} \sum_{j=1}^N e_{\nu j} =
\frac{1}{NS} \sum_{j=1}^N \; \lgl \; S_j^\nu \; \rgl
\ee
being the average over the sample $\nu$-th spin component.

Note that $j$ is the total electric current in the circuit generated by all spins inside
the coil. This is why the expression for the electromotive force contains the summation
over all spins. Respectively, the feedback magnetic field $H$, proportional to the total
electric current $j$, is the field resulting from the summary action of all spins,
independently from their location inside the coil. In practice, the distribution of magnetic
defects can be arbitrary and we do not require any translational invariance. And really we
consider the case, when the defects are located at the graphene sample edges. Their location
does not change the total current induced in the coil, but does influence the equations of
motion and, respectively, the overall dynamics. The equations of motion, treated in the
following section, study the case of magnetic defects located at the edges of graphene flakes.

Angle brackets imply statistical averaging
\be
\label{K9}
\lgl \; S_j^\al \; \rgl = {\rm Tr} \hat\rho(0) S_j^\al(t) =
\lgl \; S_j^\al(t) \; \rgl  \;   ,
\ee
in which $\hat{\rho}(0)$ is the statistical operator at the initial time $t = 0$. As far as
the system is nonequilibrium, the standard notion of temperature, strictly speaking, is not
defined. In principle, for nonequilibrium spins, it could be possible to introduce a negative
effective spin temperature dependent on time \cite{Abragam_14}. This, however, is not
necessary, since the time evolution of average spins is prescribed by the Heisenberg equations
of motion. And all information on the initial statistical operator can be included in the
initial values
\be
\label{K10}
  e_\al(0) = \frac{1}{NS} \sum_{j=1}^N \lgl \; S_j^\al(0) \; \rgl =
\frac{1}{NS} \sum_{j=1}^N {\rm Tr} \hat\rho(0) S_j^\al(0) \; ,
\ee
where the left-hand side is given as an initial condition.

The initial conditions to equation (\ref{5}) are
\be
\label{10}
h(0) = \dot{h}(0) = 0 \;   ,
\ee
where the overdot signifies time derivative. The electric circuit is tuned in
resonance with the Zeeman frequency,
\be
\label{11}
 \om_0 = \frac{1}{\hbar}\; |\; \mu_S B_0 \; | \;  ,
\ee
so that
\be
\label{12}
 \om = \om_0 \; ,
\ee
because of which the electric circuit is termed a resonance circuit.

Magnetic defects are assumed to be incorporated into a graphene flake that is in
equilibrium, hence being characterized by temperature. Generally speaking, the interaction
of magnetic defects with the graphene matrix can result in the appearance of temperature
effects in spin motion. The main influence of these effects is in the arising attenuation
in spin dynamics caused by the interaction of spins with phonon fluctuations. This leads
to the emergence of the so-called attenuation term $\gamma_1 = 1/T_1$ in the equations of
spin motion \cite{Abragam_14} that is of the order of the strength of spin-phonon interactions
and depending on temperature. However, considering short-time spin dynamics, the term
$\gamma_1$ can be safely omitted. This is because fast coherent spin reversal, as is studied
in the present paper, occurs during the time shorter than the time of spin dephasing $T_2$
being of the order of the strength of spin interactions. Spin-phonon interactions are
practically always much smaller than spin interactions, so that $T_2 \ll T_1$. This means
that if the main processes occur at times shorter than $T_2$, and  $\gamma_1 \ll \gamma_2$,
then the attenuation $\gamma_1 = 1/T_1$ can be neglected. So, the dependence of $\gamma_1$
on temperature is of no importance. Of course, the said above assumes that temperature is
not too high, being such that $k_B T / J_{ij} \ll 1$. Under this condition, spin dynamics
is much more influenced by spin interactions, but not by temperature effects. If the
nearest-neighbor spin interactions are of order $J_{ij} \sim 10^{-13}$ erg, as is mentioned
at the beginning of this section, this means that temperature is considered as low for
$T \ll 10^3$ K. Throughout the paper, we keep in mind that graphene temperature is low in
the above sense, so that temperature effects are negligible.

\section{Equations of motion for spin variables}

We write the equations for the spin variables (\ref{8}) using the Heisenberg
equations of motion
\be
\label{13}
i\hbar \; \frac{de_{\nu j}}{dt} = \frac{1}{S} \;
\lgl \; [\;  S_j^\nu , \; \hat H \; ] \; \rgl
\ee
and then calculate the time dependence of variable (\ref{9}) employing the mean-field
approximation
\be
\label{14}
 \lgl \; S_i^\mu S_j^\nu \; \rgl = \lgl \; S_i^\mu \; \rgl \lgl \; S_j^\nu \; \rgl \;  .
\ee

At the initial moment of time, spins are polarized along the axis $z$ so that
the distribution of up or down spins at their sites is random, but their average
polarization (\ref{9}) is fixed as
\be
\label{15}
 \lgl \; e_z(0) \; \rgl = - 0.9 \;  .
\ee

The external magnetic field $B_0$ is directed so that the initial spin
polarization (\ref{15}) corresponds to a nonequilibrium state, while the
equilibrium spin polarization would  correspond to the up direction. Our aim
is to find the system parameters that could produce fast spin reversal, when
the reversal time, that is, the time from the starting point to the moment of
time, where the average spin is reversed, would be short. Also, it is desirable
that the reversal be practically complete and after the reversal the average
spin would not strongly fluctuate. Such a regime of spin reversal is optimal
for applications in spintronics and information processing.

\section{Defects at zigzag edges of graphene flakes}

As a concrete system, we consider magnetic defects caused by hydrogenation of
zigzag edges of graphene flakes. Spins interact through the nearest neighbors,
with a ferromagnetic exchange potential $J > 0$. For simplicity, the value of the
external field $B_0$ is taken such that  $JS=\hbar\om_0$. We study the dynamics
of spin reversals by varying the anisotropy parameter $\alpha$, coupling parameter
$\beta$, and the resonator quality factor $Q$. We pay attention to the following
three features: (i) when the reversal time is shorter, (ii) if the reversal is
complete, and (iii) if there appear spin oscillations after the reversal.

In Fig. 1, there is no anisotropy, hence $\alpha = 1$, and the coupling parameter
is set to $\beta = 0.1$, while the quality factor is varied. The temporal behavior
of the spin polarization $e_z$ as a function of time in units of $1/\omega$ is shown.
The larger the quality factor $Q$, the shorter the reversal time. The reversal
is complete, but there appear oscillations after the reversal, if $Q > 1$.

\begin{figure}
\centerline{
\includegraphics[width=12cm]{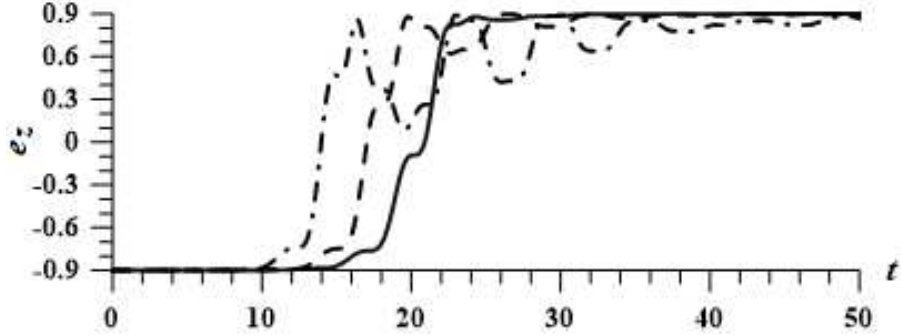} }
\caption{Longitudinal, $e_z$, spin polarization as a function of dimensionless time
(measured in units of $1/\om$), for  the coupling parameter $\bt=0.1$ and anisotropy
parameter $\al=1$ (no anisotropy) for different  resonator quality factors $Q$:
(a) $Q=1$ (solid line), (b) $Q=2$ (dashed line), and (c) $Q=5$ (dashed-dotted line). }
\label{fig:f01}
\end{figure}

The anisotropy is also absent in Fig. 2, where $\al=1$. The coupling parameter
is smaller than in the previous figure, being $\beta=0.01$. The smaller coupling
parameter makes the reversal time longer. The reversal is complete. The
after-reversal oscillations are practically absent even for $Q = 5$.

\begin{figure}
\centerline{
\includegraphics[width=12cm]{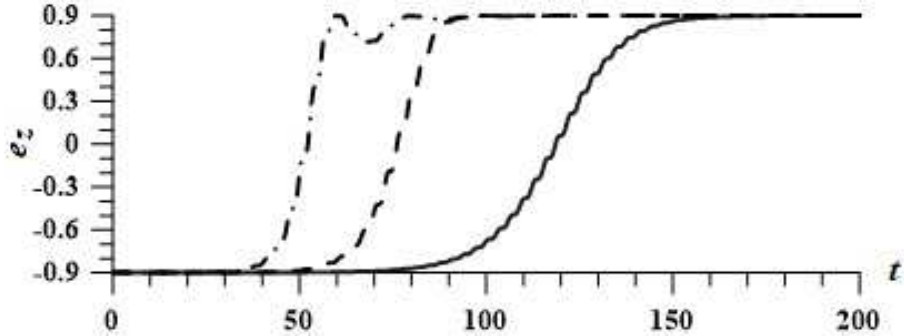} }
\caption{Longitudinal, $e_z$, spin polarization as a function of dimensionless time
(measured in units of $1/\om$), for  the coupling parameter $\bt=0.01$ and anisotropy
parameter $\al=1$ (no anisotropy) for different  resonator quality factors $Q$:
(a) $Q=1$ (solid line), (b) $Q=2$ (dashed line), and (c) $Q=5$ (dashed-dotted line). }
\label{fig:f02}
\end{figure}

Figure 3 studies the influence of the anisotropy, under the fixed coupling
parameter $\bt=0.01$ and the quality factor $Q=1$. As is expected, the larger
the anisotropy, the longer the reversal time. The parasitic oscillations are
absent, but the reversal is not complete, when the anisotropy parameter
$\al>1.5$.

\begin{figure}
\centerline{
\includegraphics[width=12cm]{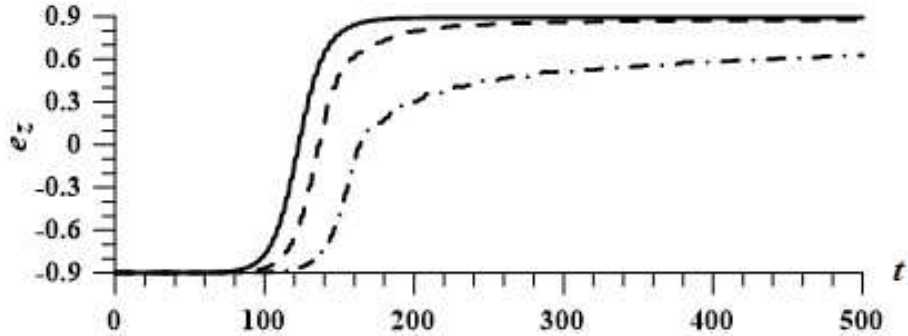} }
\caption{Longitudinal,  $e_z$, spin polarization as a function of dimensionless
time (measured in units of $1/\om$), for  the coupling parameter $\bt=0.01$ and
the resonator quality factor $Q=1$  for different anisotropy parameters $\al$:
(a) $\al=1.2$ (solid line), (b) $\al=1.4$ (dashed line), and (c) $\al=1.6$
(dashed-dotted line). }
\label{fig:f03}
\end{figure}

In Fig. 4, the role of the coupling parameter is explored, under the given $\al=1$
and $Q=1$. Increasing $\bt$ diminishes the reversal time. Slight oscillations occur
for $\beta > 0.2$.

\begin{figure}
\centerline{
\includegraphics[width=12cm]{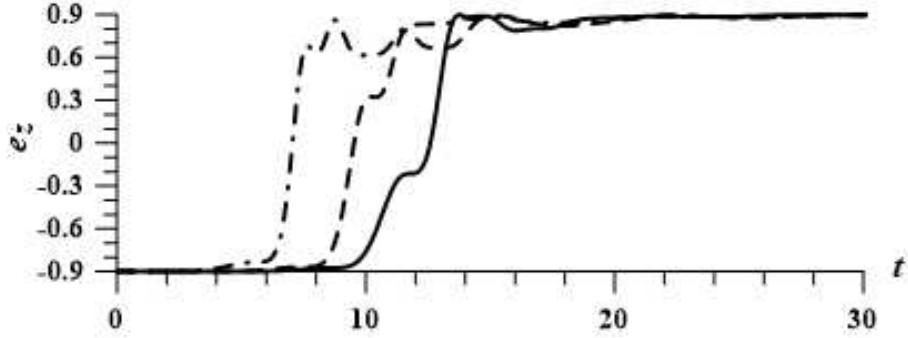} }
\caption{Longitudinal, $e_z$, spin polarization as a function of dimensionless
time (measured in units of $1/\om$), for the resonator quality factor $Q=1$
and anisotropy parameter $\al=1$, for different coupling parameters $\bt$:
(a) $\bt=0.2$ (solid line), (b) $\bt=0.3$ (dashed line), and (c) $\bt=0.4$
(dashed-dotted line). }
\label{fig:f04}
\end{figure}

Figure 5 demonstrates that increasing the quality factor, under $\alpha = 1$ and
the larger coupling parameter $\beta = 0.4$, shortens the reversal time, although
rather strong oscillations arise for $Q > 1$.

\begin{figure}
\centerline{
\includegraphics[width=12cm]{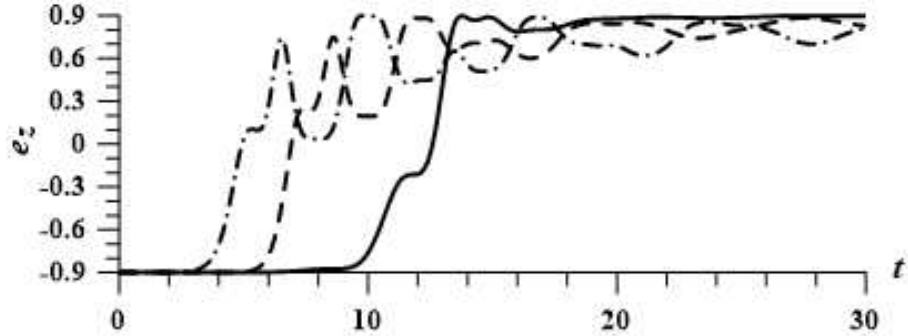} }
\caption{Longitudinal, $e_z$, spin polarization as a function of dimensionless
time (measured in units of $1/\om$), for the coupling parameter $\bt=0.4$ and
anisotropy parameter $\al=1$ for different resonator quality factors $Q$:
(a)  $Q=1$ (solid line), (b) $Q=2$  (dashed line), and (c) $Q=3$
(dashed-dotted line).}
\label{fig:f05}
\end{figure}

For sufficiently large coupling parameter, such as in Fig. 6, where $\bt=0.2$, the
reversal time does not essentially depends on the anisotropy, however under strong
anisotropy $\alpha > 1.5$ the reversal is never complete.

\begin{figure}
\centerline{
\includegraphics[width=12cm]{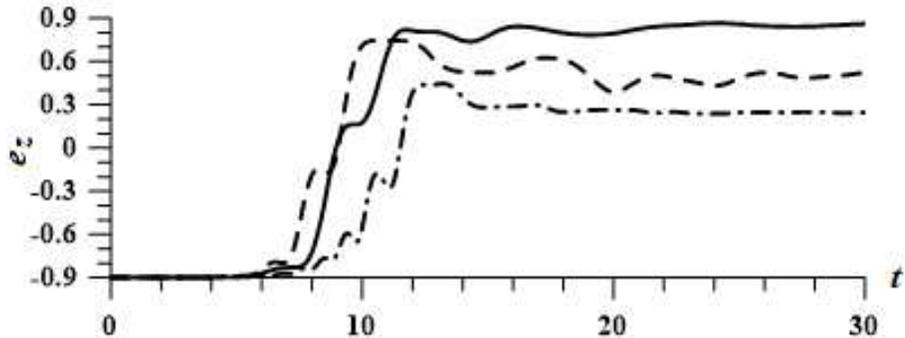} }
\caption{Longitudinal, $e_z$, spin polarization as a function of dimensionless
time (measured in units of $1/\om$), for the coupling parameter $\bt=0.2$ and
the resonator quality factor $Q=1$, for different anisotropy parameters $\al$:
(a) $\al=1.5$  (solid line), (b)  $\al=3$ (dashed line), (c) $\al=5$
(dashed-dotted line).}
\label{fig:f06}
\end{figure}

In that way, to realize optimal conditions for spin reversal, requiring the
validity of three criteria, short reversal time, complete spin reversal, and the
absence of parasitic oscillations, one has to appropriately adjust the system
parameters $\alpha$, $\beta$, and $Q$. From numerical calculations, we find for
the product of these parameters the condition $\alpha \beta Q \sim 0.1$, when all
three criteria are satisfied. Keeping in mind the definition of $\beta$ and $Q$,
we come to the condition
\be
\label{16}
 \frac{\rho\mu_S^2 S}{\hbar\gm} \sim 0.1 \;  ,
\ee
where $\rho$ is the density of magnetic defects. In this way, for realizing
optimal spin reversal in graphene flakes, one needs to choose the system
parameters satisfying estimate (\ref{16}).

\section{Conclusion}

We have described a method of regulating spin dynamics of the so-called magnetic
graphene. Magnetic moments can be incorporated into graphene flakes and graphene
ribbons through defects, such as adatoms and vacancies. Also, spins can be located
at hydrogenated zigzag edges of graphene. Spin reversal can be produced by
connecting the sample to a resonance electric circuit. The action of the resonator
feedback field strongly accelerates spin reversal. This method is essentially simpler
than spin reversal by means of transverse alternating fields, where one needs to define
the field amplitude, its frequency, and duration. The feedback field adjusts automatically,
being created by moving spins themselves. Choosing the needed parameters, one can easily
regulate the spin reversal time, finding the conditions when there are no after-reversal
oscillations. The described method of fast and easily regulated spin reversal can find
applications in spintronics and in quantum information processing. Other applications
can be connected with the electromagnetic radiation emitted by coherently moving spins
\cite{Yukalov_14}, as it is generated in different systems under coherent spin motion
\cite{Belozerova_15,Yukalov_16,Yukalov_17,Yukalov_18,Yukalov_19,Yukalov_20,Yukalov_21}.
It is important to stress that the coherent spin motion, when no strong external
coherent fields are involved, necessarily requires the presence of a resonator
\cite{Yukalov_22}.

\section*{Acknowledgement}

One of the authors (V.I.Y.) is grateful to E.P. Yukalova for discussions.

\end{document}